\documentclass[referee]{aa}

\usepackage{graphicx}
\usepackage{natbib}
\usepackage{latexsym}
\usepackage{amsmath}
\usepackage{amssymb}
\usepackage{amstext}
\usepackage{color}
\usepackage{psfrag}
\usepackage{txfonts}

\def\k{km s$^{-1}$}
\def\ks{km s$^{-1}$~}

\def\m{$^\prime$}
\def\s{$^{\prime\prime}$}

\def\cm3{cm$^{-3}$}

\def\2{$^{12}$CO}
\def\3{$^{13}$CO}
\def\msol{M$_\odot$}

\begin{document}

\title{The HII region G35.673-00.847: another case of triggered star formation?}
\author {S. Paron \inst{1,2}
\and A. Petriella \inst{1}
\and M. E. Ortega  \inst{1}
}

\institute{Instituto de Astronom\'{\i}a y F\'{\i}sica del Espacio (IAFE),
             CC 67, Suc. 28, 1428 Buenos Aires, Argentina\\
             \email{sparon@iafe.uba.ar}
\and FADU - Universidad de Buenos Aires
}
\offprints{S. Paron}

   \date{Received <date>; Accepted <date>}

\abstract{}{As part of a systematic study that we are performing with the aim to increase
the observational evidence of triggered star formation in the surroundings of HII regions, we analyze 
the ISM around the HII region G35.673-00.847, a poorly studied source.}{Using data from large-scale surveys:
Two Micron All Sky Survey, Galactic Legacy Infrared Mid-Plane Survey Extraordinaire (GLIMPSE), MIPSGAL, 
Galactic Ring Survey (GRS), VLA Galactic Plane Survey (VGPS), and NRAO VLA Sky Survey (NVSS) we performed a 
multiwavelength study of G35.673-00.847 and its surroundings.}{The mid IR emission, shows that G35.673-00.847 
has an almost semi-ring like shape with a cut towards the galactic west. The radius of this semi-ring is about 1\farcm5 
($\sim 1.6$ pc, at the distance of $\sim 3.7$ kpc). 
The distance was estimated from an HI absorption study and from the analysis of the molecular gas. 
Indeed, we find a molecular shell composed by several clumps distributed around the HII region, suggesting that 
its expansion is collecting 
the surrounding material. We find several YSO candidates over the molecular shell.
Finally, comparing the HII region 
dynamical age and the fragmentation time of the molecular shell, we discard the collect and collapse as the 
mechanism responsible for the YSOs formation, suggesting other processes such as radiative driven implosion 
and/or small-scale Jeans gravitational instabilities. }{}

\titlerunning{The HII region G35.673-00.847}
\authorrunning{S. Paron et al.}


\maketitle

\section{Introduction}

During the last years, the Galactic Legacy Infrared Mid-Plane Survey Extraordinaire (GLIMPSE) performed with data 
obtained from the {\it Spitzer Space Telescope}, was a very useful tool to study the Galactic IR emission
with unprecedented quality and resolution, and still remains so. Using these mid-IR data, for instance, 
it is possible to clearly identify the photodissociation regions (PDRs) surrounding HII regions. Thus, from a 
multiwavelength analysis can be 
studied the interaction between the HII region and the surrounding interstellar medium (ISM), and eventually discover 
triggered star formation. 
One of the triggered processes recently largely studied in the HII region borders is the ``collect and collapse'', 
which was early proposed by \citet{elme77}.
In such process, during the supersonic expansion of an HII region, a dense layer of material can be collected 
between the ionization and the shock fronts. This layer can be fragmented in 
massive condensations that then can collapse to lead the formation of new massive stars and/or clusters. 
Recent observational studies support that this mechanism is taking place in several HII regions 
(see e.g. \citealt{alberto10,poma09,zav07}, and references therein).

G35.673-00.847 (hereafter G35.6) is an HII region poorly studied. The source was cataloged in the HII region 
catalogue of \citet{lock89}, who obtained a recombination line at v$_{\rm LSR} \sim 60$ \k. According to the 
IRAS catalogue of Point Sources, G35.6 coincides with the source IRAS 18569+0159. 
In the NRAO VLA Sky Survey (NVSS), \citet{condon98} 
identified two radio sources: NVSS 185929+020334~and 185938+020012~towards this region. 

This work is part of a systematic study that we are performing with the aim to increase
the observational evidence of triggered star formation in the surroundings of HII regions. 
We present a molecular and near- and mid-IR study of the environment that surrounds
the HII region G35.6~with the purpose of exploring the ISM around it, and looking for signatures of 
star formation.

\section{Data}

We analyzed data extracted from four large-scale surveys:
Two Micron All Sky Survey (2MASS)\footnote{2MASS is a joint project of
the University of Massachusetts and the Infrared Processing and Analysis Center/California Institute of Technology,
funded by the National Aeronautics and Space Administration and the National Science Foundation.},
Galactic Legacy Infrared Mid-Plane Survey Extraordinaire (GLIMPSE),
MIPSGAL and GRS\footnote{Galactic Ring Survey \citep{jackson06}}. 

GLIMPSE is a mid infrared survey of the inner 
Galaxy performed using the {\it Spitzer Space Telescope}. We used the mosaicked images from
GLIMPSE and the GLIMPSE Point-Source Catalog (GPSC) in the {\it Spitzer}-IRAC (3.6, 4.5, 5.8 and 8 $\mu$m).
IRAC has an angular resolution between 1\farcs5 and 1\farcs9 (see \citealt{fazio04} and \citealt{werner04}).
MIPSGAL is a survey of the same region as GLIMPSE, using MIPS instrument (24 and 70 $\mu$m) on {\it Spitzer}.
The MIPSGAL resolution at 24 $\mu$m is 6\s. 

The GRS was performed by the Boston University and the
Five College Radio Astronomy Observatory (FCRAO). The survey maps the Galactic Ring in the \3 J=1--0 line
with an angular and spectral resolution of 46\s~and 0.2 \k, respectively (see \citealt{jackson06}).
The observations were performed in both position-switching and On-The-Fly mapping modes, achieving an
angular sampling of 22\s. 

Additionally we used HI data with an angular resolution of $\sim$1\m~extracted from the VLA Galactic Plane 
Survey (VGPS; \citealt{stil06}) and radio continuum data extracted from the NRAO VLA Sky Survey (NVSS) with 
an angular resolution of $\sim$45\s~\citep{condon98}.

\section{Presentation of G35.673-00.847 (G35.6)}

\begin{figure*}[tt]
\includegraphics[width=6.8cm]{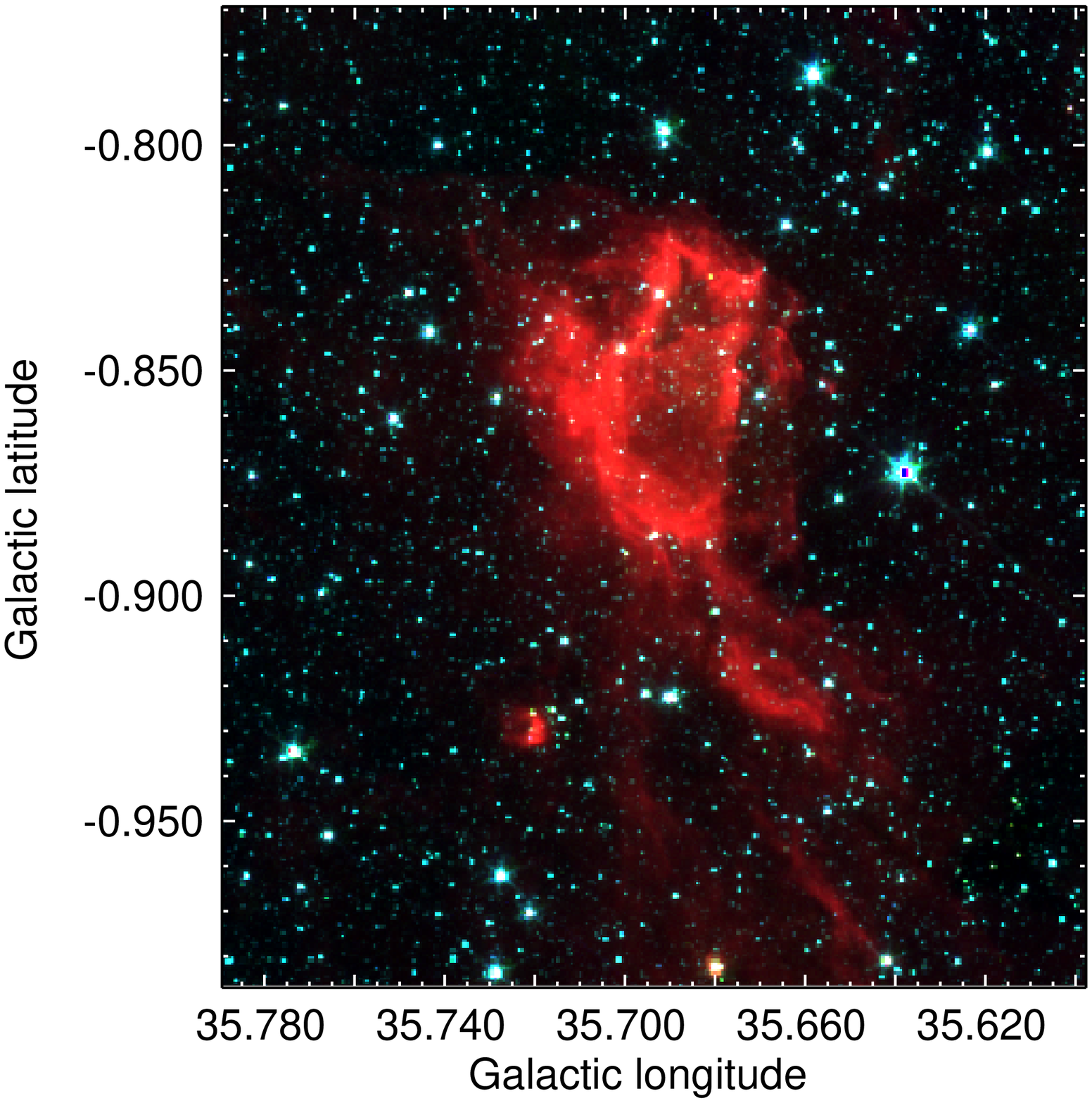}
\includegraphics[width=7cm]{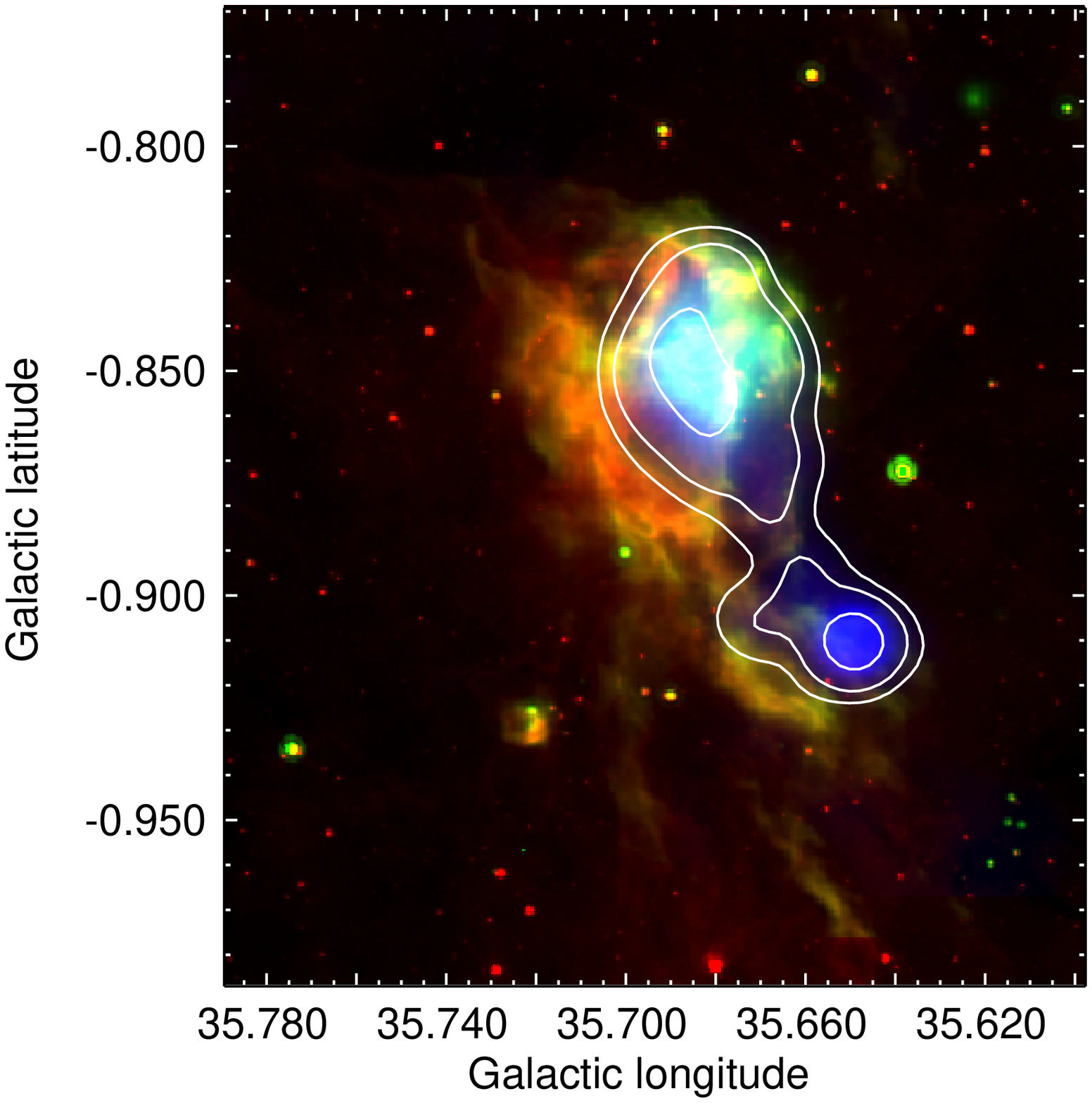}
\caption{Left: {\it Spitzer}-IRAC three color image 
(3.6 $\mu$m $=$ blue, 4.5 $\mu$m $=$ green and 8 $\mu$m $=$ red). Right: color composite image where 
the {\it Spitzer}-IRAC 8 $\mu$m emission is displayed in red, the {\it Spitzer}-MIPSGAL emission at 24 $\mu$m in green, 
and the NVSS radio continuum emission at 20 cm is presented in blue and emphasized with white contours with levels of 
2.5, 6 and 20 mJy beam$^{-1}$. The $\sigma_{\rm rms}$ of the NVSS data is 0.45 mJy beam$^{-1}$.} 
\label{figpresent}
\end{figure*}

Figure \ref{figpresent} shows two composite three color images of G35.6. The left image displays
three {\it Spitzer}-IRAC bands:  3.6 $\mu$m (in blue), 4.5 $\mu$m (in green) and 8 $\mu$m (in red). 
The right image shows the {\it Spitzer}-IRAC emission at 8 $\mu$m (in red), the {\it Spitzer}-MIPSGAL emission 
at 24 $\mu$m (in green) and the NVSS radio continuum emission at 20 cm (in blue and emphasized with 
white contours). 
Both figures clearly show the PDR visible in the 8 $\mu$m emission, which is 
mainly originated in the polycyclic aromatic hydrocarbons (PAHs). The PAHs emission delineates the HII region 
boundaries because these large molecules are destroyed inside the ionized region, but are excited in the 
PDR by the radiation leaking from the HII region \citep{poma09}.
The 24 $\mu$m emission reveals the presence of hot dust, and the radio continuum emission shows two sources, one
related to G35.6, probably due to its ionized gas, and other lying towards the south. Both 
sources are: NVSS 185929+020334 and 185938+020012, respectively \citep{condon98}.
The PAH emission shows that G35.6~has an almost semi-ring like shape with a cut towards the galactic west. 
The radius of this semi-ring is about 1\farcm5. Extending towards 
the south is visible another PDR, which can be related to the radio continuum source NVSS 185938+020012.
On the other hand, a little bubble is present in the field, at $l = 35$\fdg722, $b = -0$\fdg928. The
emission at 8 $\mu$m and 24 $\mu$m, showing the presence of PAH and hot dust towards this bubble, suggest that it 
could be a young HII region.

\section{Distance}

\begin{figure}
\centering
\includegraphics[totalheight=0.27\textheight,angle=-90,viewport=0 0 566 730,clip]{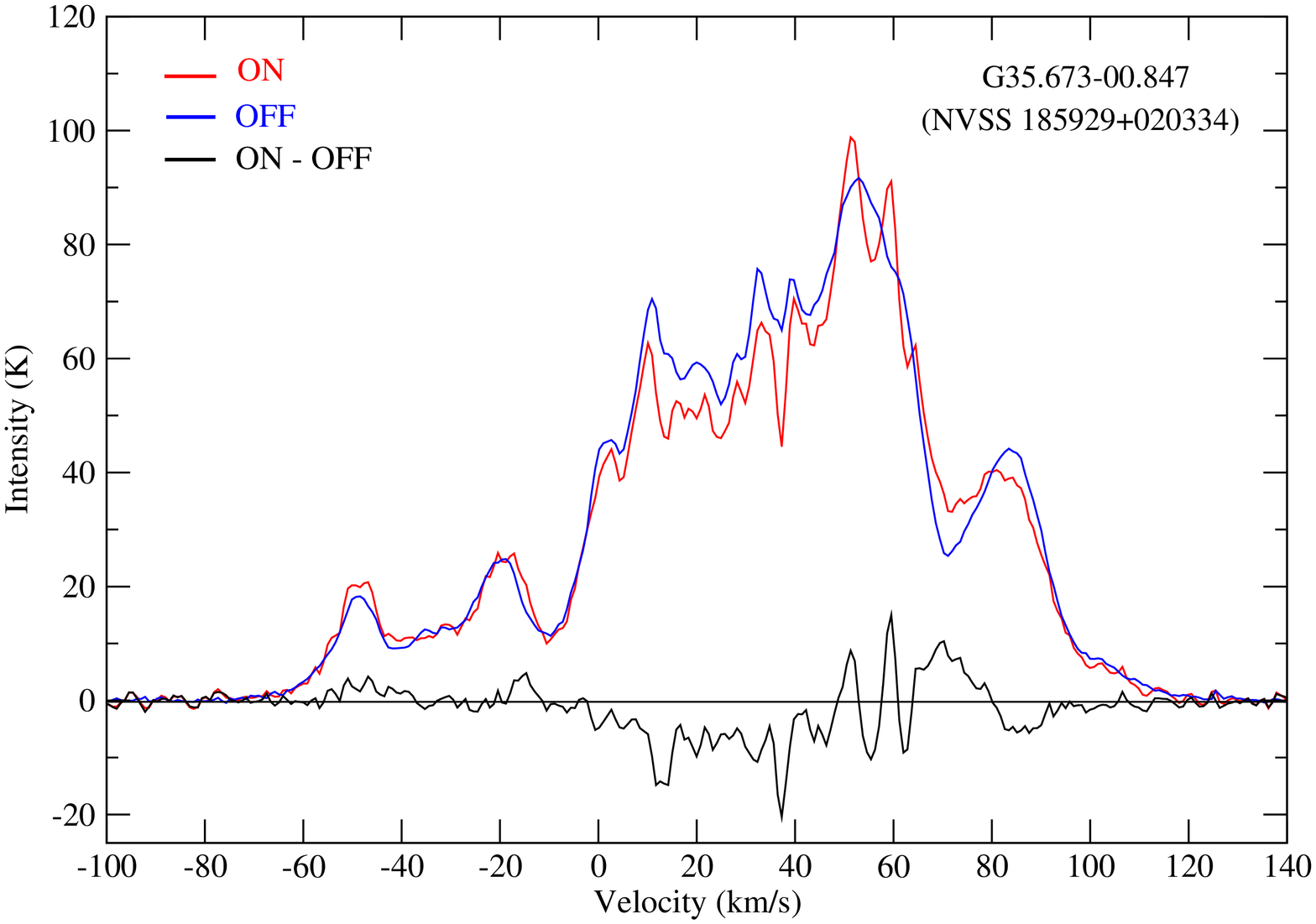}
\includegraphics[totalheight=0.27\textheight,angle=-90,viewport=0 0 566 730,clip]{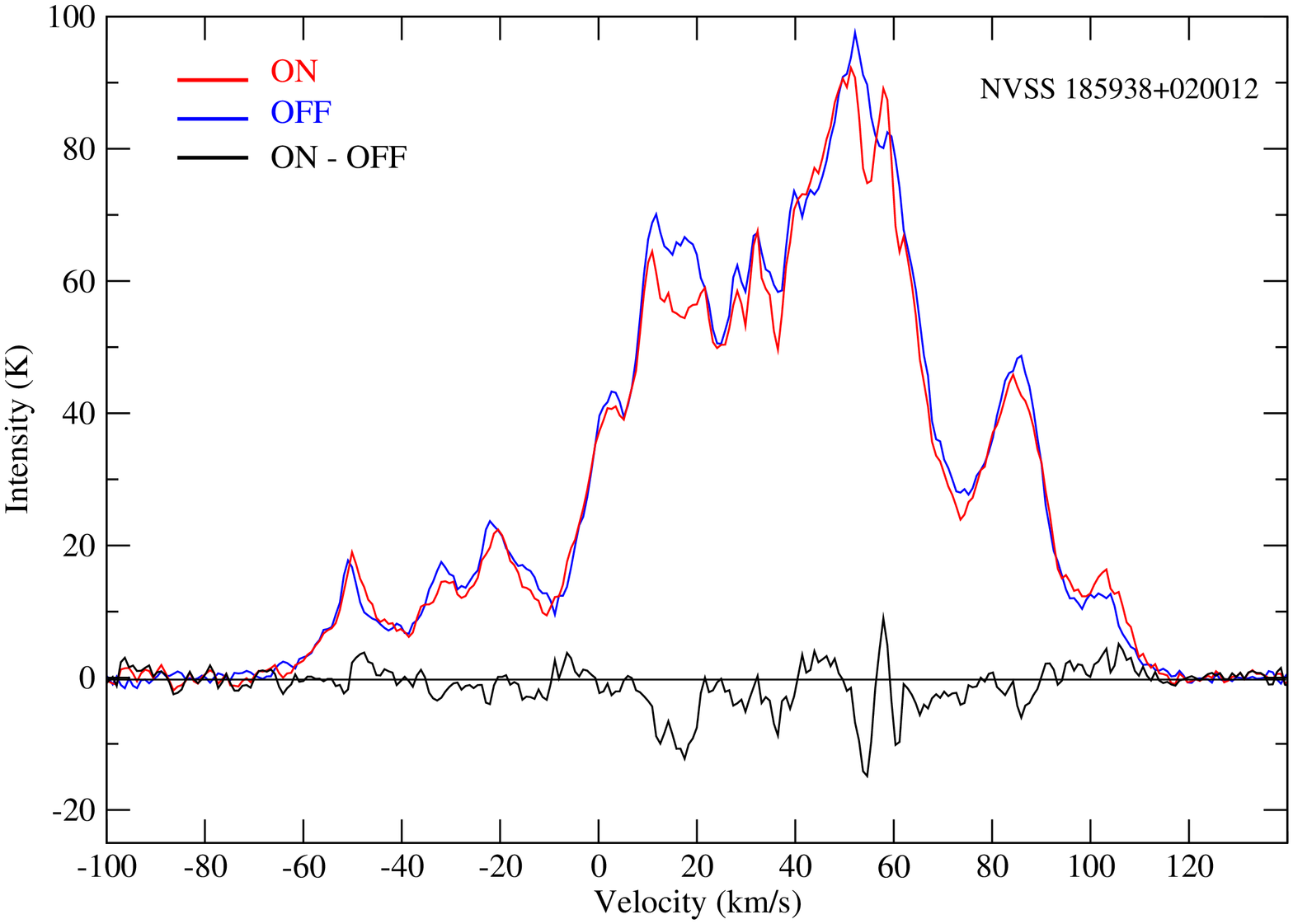}
\caption{Left: HI spectra obtained towards the source G35.6~(NVSS 185929+020334). Right: HI spectra
obtained towards the source NVSS 185938+020012. The spectra obtained towards the sources 
(the On position) are presented in red, in blue is presented the averaged HI emission  
taken from four positions separated by approximately a beam from the source in direction of the four galactic 
cardinal points (the Off position), and the subtractions between them are presented in black. The 3$\sigma$ 
uncertainty of the subtraction is $\sim$10 K.} 
\label{hiabs}
\end{figure}

G35.6~presents a radio recombination line at v$_{\rm LSR} \sim 60$ \ks \citep{lock89}, 
which, by applying the flat galactic rotation curve of \citet{Fich89}, that assumes circular rotation around 
the galactic center, gives the possible kinematic distances 
of $\sim 4.0$ or $\sim 9.8$ kpc. This ambiguity arises because we are studying a region in the first galactic 
quadrant, where a given velocity may be associated with two possible distances. 
Using HI data we performed an absorption study towards the radio sources G35.6~(NVSS 185929+020334)
and NVSS 185938+020012.
Figure \ref{hiabs} shows the HI spectra towards both sources. 
The HI emission obtained over the source (the On position: a beam
over the radio maximum of the source) is presented in red, in blue is presented the average HI emission taken 
from four positions 
separated by approximately a beam from the source in direction of the four galactic cardinal points (the Off position),
and the subtraction between them is presented in black, which has a 3$\sigma$ uncertainty of $\sim$10 K.
The figure shows that both sources have similar HI 
absorption features, suggesting that they are located at the same distance. The last absorption feature appears
at v $\sim$ 61 \k, in coincidence with the G35.6 recombination line \citep{lock89}. Taking into account 
that the tangent
point (at v $\sim$ 89.7 \k) does not present any absorption, following \citet{kolpak03}, we favour the near 
kinematic distance. 

\section{Molecular analysis}
\label{secmolec}

\begin{figure}[h]
\centering
\includegraphics[width=10cm]{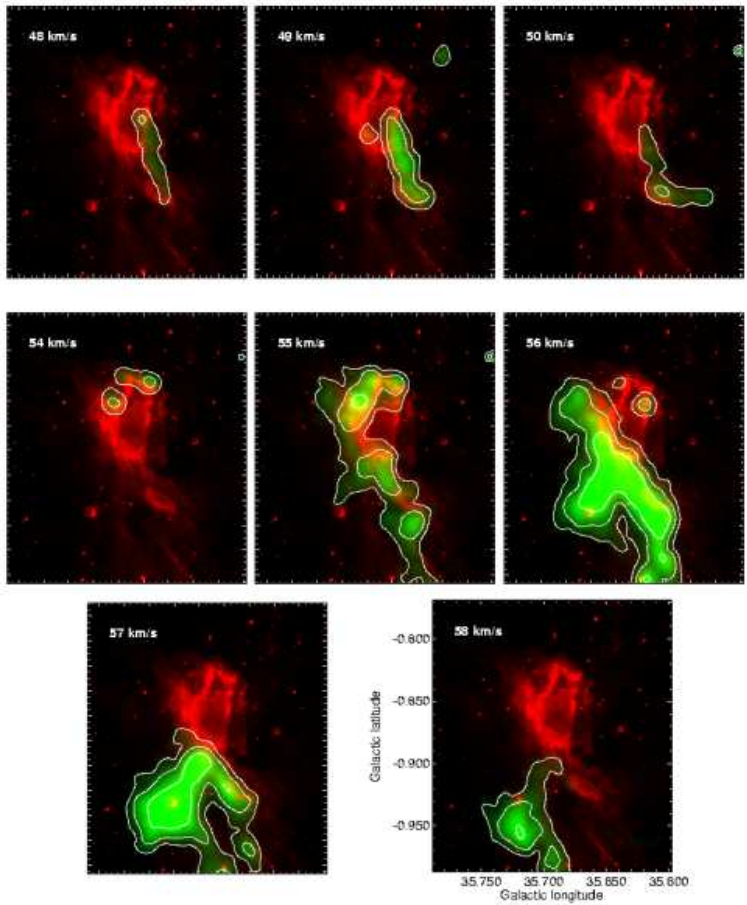}
\caption{Integrated velocity channel maps of the \3 J=1--0 emission (in green) every $\sim 1$ \ks in two velocity
intervals: from $\sim 48$ to 51 \ks (shown in the first three panels) and from $\sim 54$ to 59 \ks (shown in the 
remaining panels). The contour levels of the \3 J=1--0 emission are 1, 2 and 4 K \k. Red is the 8 $\mu$m emission.}
\label{panel}
\end{figure}

\begin{figure}[h]
\centering
\includegraphics[width=6.5cm]{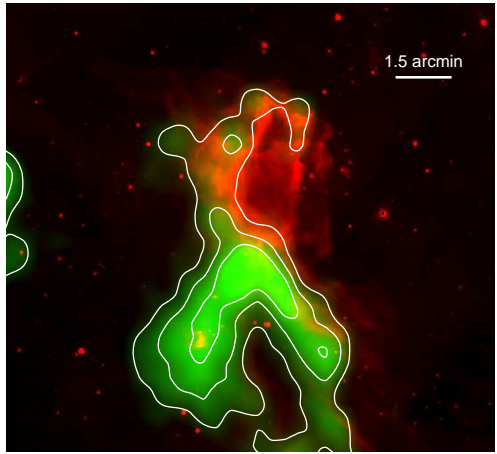}
\caption{\3 J=1--0 emission (in green) integrated between 53 and 61 \k. The contour levels of the \3 J=1--0 emission 
are 4.5, 8 and 12 K \k. Red is the 8 $\mu$m emission.}
\label{integ}
\end{figure}

We analyze the whole \3 J=1--0 data cube and we find some interesting molecular structures between 47 and 60 \k. 
Figure \ref{panel} displays the integrated velocity channel maps of the \3 J=1--0 emission every $\sim$ 1 \k, showing
the kinematical and morphological structure of a molecular cloud probably related to the HII region G35.6. 
Between $\sim 48$ and 51 \ks appears a molecular structure delineating the PDR that extends to the south. 
No molecular gas is observed between 51 and 54 \ks (this velocity interval is not shown in Fig. \ref{panel}).
Finally, between 54 and 59 \ks appear several molecular clumps distributed over the 
borders of G35.6 and the southern PDR, which may indicate that the collect and collapse 
process could be taking place in this region. As \citet{deha05} point out, the presence of a dense
molecular shell surrounding the ionized gas of an HII region, or the presence of massive fragments regularly
spaced along the ionization front, can prove that we are dealing with the collect and collapse mechanism.
Figure \ref{integ} shows the \3 J=1--0 emission integrated between 53 and 61 \ks (in green) over the 8 $\mu$m emission
(in red). The very good correspondence between the eastern HII region border, traced by the IR emission, and the molecular 
gas, strongly suggests that the observed molecular shell has been swept and shaped by the expansion of G35.6. 
The central velocity of the molecular gas is $\sim 57$ \k, which gives the 
possible kinematic distances of 3.7 or 10.1 kpc. According to the study presented in Sec. 4, we 
favour the nearest one. 
Taking into account that the ionized gas may be moving away from the molecular material, we use the central velocity
of the molecular gas to adopt 3.7 kpc as the distance of the whole complex.

In order to have an estimate of the mass and density of the described molecular shell, we 
assume LTE, an excitation temperature of 20 K, a distance of 3.7 kpc, and that the
\3 emission is optically thin. From the standard LTE equations, 
we obtain a N(\3) $\sim 3 \times 10^{16}$ cm$^{-2}$,
and using the relation N(H$_{2}$)/N(\3)$ \sim 5 \times 10^5$ (e.g. \citealt{simon01}), we
obtain a molecular mass and a density of $\sim 1.5 \times 10^{4}$ \msol~and $\sim 1 \times 10^{4}$ cm$^{-3}$, 
respectively. The integration was performed over all the observed
positions within the 4.5 K \ks contour level shown in Fig. \ref{integ}, following the shell geometry 
shown at the 55 \ks channel map in Fig. \ref{panel}, i.e., the molecular 
condensation extending towards the southeast was not considered. To calculate the volume of the 
molecular shell, we assume a length along the line of sight of $\sim$1\m~($\sim$ 1.1 pc at the distance
of 3.7 kpc), which is approximately the average of the shell width seen in the plane of the sky 
(see the 55 \ks channel map in Fig. \ref{panel}).
On the other hand, it is important to 
note that the little bubble described in Sec. 3, probably a young HII region, is likely embedded in 
this molecular condensation, suggesting to be active in star formation.

\section{Exciting stars}
\label{excit}

No exciting star of the HII region G35.6 was found in the
literature. In this work, we give some indirect evidence suggesting the
possible location and properties of the exciting star(s) of the
region.

The first piece of information is given by the radio continuum
emission of G35.6, which allow us to derive the expected spectral type
of the exciting star.  The number of UV ionizing photons needed to
keep an HII region ionized is given by $N_{\mathrm{uv}}= 0.76 \,
\times 10^{47}\, T_4^{-0.45}\, \nu_{\rm GHz}^{0.1}\,S_{\nu}\, D_{\rm
kpc}^2 $~\citep{cha76}, where $T_{\mathrm{4}}$ is the electron
temperature in units of $10^4$~K, $D_{\rm kpc}$ the distance in kpc,
$\nu_{\rm GHz}$ the frequency in GHz, and $S_{\nu}$ the measured total
flux density in Jy.  Assuming an electron temperature of T $= 10^{4}$
K, a distance of 3.7 kpc, and using a total flux density of 0.86 Jy at
2.7 GHz for G35.6 \citep{reich84} and a total flux density of 0.053 Jy at
1.4 GHz for NVSS 185938+020012 \citep{condon98}, the total amount of ionizing photons
needed to keep these sources ionized turns out to be
about $N_{\mathrm{uv}} = 1.0 \times 10^{48}$~ph~s$^{-1}$ and
$N_{\mathrm{uv}} = 0.6 \times 10^{47}$~ph~s$^{-1}$, respectively.
It is well established that part of the UV radiation can be dissipated
in heating the dust. In fact, \citet{ino01I} and \citet{ino01II}
demonstrated that typically only half of the Lyman continuum photons
from the central source in a Galactic HII region ionizes neutral
hydrogen, and the rest are absorbed by dust grains within the ionized
region. Taking this into account, considering errors of about ten
percent in the distance and in the radio continuum flux at 2.7 GHz, and
based on the ionizing fluxes for massive stars given by
\citet{martins05}, the estimated spectral type of the ionizing star of
G35.6 ranges between O7.5V and O9V.

\begin{figure}[h]
\centering
\includegraphics[width=8.5cm]{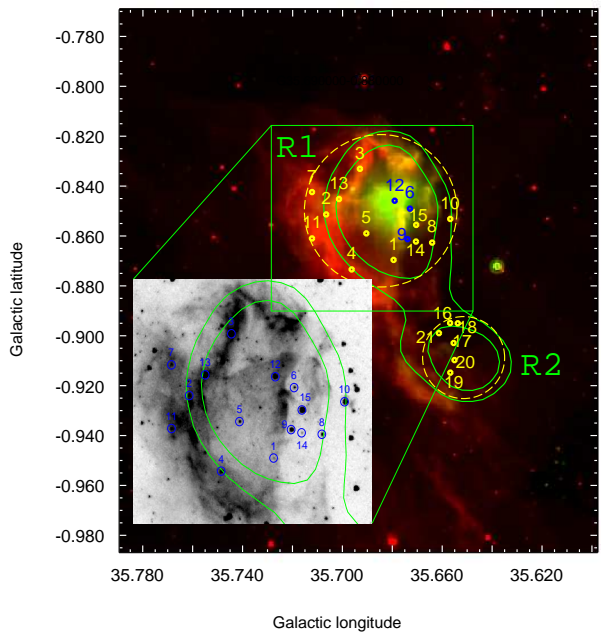}
\caption{{\it Spitzer}-IRAC two color image (8 $\mu$m = red and 24
$\mu$m = green). The green contours represent the radio continuum
emission at 20 cm. The crosses show the location of the main
sequence star candidates in R1 and R2 (dashed circles).}
\label{Ostars}
\end{figure}

\begin{table}
\caption{Exciting star candidates in regions R1 and R2. The sources number correspond to the numeration
in Fig. \ref{Ostars}.}
\label{MSSC}
\tiny
\centering
\begin{tabular}{cccccccccc}
\hline\hline
\#  & 2MASS Designation & $J$ & $H$ & $K$ &  $A_v$ & M$_J$ & M$_H$ & M$_K$ & O-type star\\
\hline
& & &  & & {\it R1}&  \\
1   &  J18593297+0202530  & 11.92 & 11.52 & 11.40 &   4.18 & -2.10 & -2.05 & -1.91 &  -\\
2   &  J18593204+0204495  & 16.67 & 13.34 & 11.69 &  30.26 & -4.70 & -4.80 & -4.54 & {\it yes}\\
3   &  J18592664+0204362  & 13.90 & 10.54 &  8.86 &  30.61 & -7.58 & -7.66 & -7.41 &  -\\
4   &  J18593563+0203404  & 15.05 & 12.25 & 10.88 &  25.44 & -4.97 & -5.05 & -4.81 & {\it yes}\\
5   &  J18593189+0203454  & 14.66 & 12.11 & 11.01 &  22.12 & -4.42 & -4.60 & -4.31 & {\it yes}\\
6   &  J18592786+0203057  & 14.14 & 11.88 & 10.92 &  19.58 & -4.23 & -4.39 & -4.11 & {\it yes}\\
7   &  J18593076+0205225  & 11.11 & 10.60 & 10.52 &   4.31 & -2.95 & -2.99 & -2.80 &  -\\
8   &  J18592979+0202149  & 13.35 & 11.05 & 10.05 &  20.17 & -5.18 & -5.32 & -5.05 &  -\\
9   &  J18593057+0202486  & 13.28 & 10.70 &  9.53 &  22.81 & -6.00 & -6.13 & -5.87 &  -\\
10  &  J18592697+0202073  & 12.66 & 10.11 &  8.89 &  23.09 & -6.69 & -6.77 & -6.53 &  -\\
11  &  J18593472+0204518  & 16.34 & 13.32 & 11.88 &  27.07 & -4.13 & -4.26 & -3.99 & {\it yes}\\
12  &  J18592785+0203304  & 11.69 & 10.30 &  9.56 &  13.83 & -5.05 & -4.96 & -4.83 & {\it yes}\\
13  &  J18593016+0204431  & 13.84 & 10.65 &  8.90 &  30.39 & -7.57 & -7.51 & -7.35 &  -\\
14  &  J18593041+0202363  & 14.38 & 12.47 & 11.64 &  16.95 & -3.24 & -3.34 & -3.10 & {\it yes}\\
15  &  J18592897+0202470  & 11.54 &  9.09 &  7.87 &  22.60 & -7.67 & -7.71 & -7.50 &  -\\
\hline
& & &  & & {\it R2} &\\
16  &  J18593584+0200579  & 14.35 & 12.37 & 11.51 & 17.50  & -3.43 & -3.53 & -3.29 & {\it yes}  \\
17  &  J18593743+0200411  & 14.07 & 11.69 & 10.58 & 21.40  & -4.81 & -4.90 & -4.66 & {\it yes}   \\
18  &  J18593556+0200488  & 13.81 & 11.97 & 11.25 & 15.70  & -3.46 & -3.62 & -3.35 & {\it yes}    \\
19  &  J18594015+0200255  & 15.76 & 12.94 & 11.67 & 24.73  & -4.06 & -4.23 & -3.94 & {\it yes}  \\
20  &  J18593887+0200288  & 15.20 & 12.82 & 11.71 & 21.40  & -3.68 & -3.77 & -3.53 & {\it yes}  \\
21  &  J18593725+0201066  & 15.22 & 13.21 & 12.25 & 18.47  & -2.83 & -2.86 & -2.66 &  -  \\
\hline
\end{tabular}
\end{table}

Additionally we perform a photometric study of the infrared point sources in the region based on the
GLIMPSE I Spring'07 and the 2MASS All-Sky Point Source Catalogs. Only sources with
detections in the four {\it Spitzer}-IRAC bands and in the three 2MASS
bands were considered. We find 30 and 8 sources towards the HII region
G35.6 (R1 in Fig. \ref{Ostars}) and towards the source NVSS
185938+020012 (R2 in Fig. \ref{Ostars}), respectively.
To examine the evolutionary stage of the infrared point sources, we
analyze their location onto a color-color IRAC diagram. Following
\citet{all04} color criteria, we found 15 and 6 sources in R1 and R2,
respectively, that can be classified as main sequence stars (Class
III). Table \ref{MSSC} presents these sources with its 2MASS
designation (Col. 2), apparent {\it JHK} magnitudes (Cols. 3-5),
estimated extinctions (Col. 6), calculated absolute {\it JHK}
magnitudes (Cols. 7-9) and when its derived spectral type coincides with an 
O-type star is remarked in Col. 10. The errors in the estimated extinctions and 
in the calculated absolute {\it JHK} magnitudes are below 20\% and 30\%, respetively.
The sources are labeled according to
Fig. \ref{Ostars}, which displays their location into a two color
image, where the 8~$\mu$m and 24~$\mu$m emissions are displayed in red and green, respectively.  
The green contours delineate the radio continuum emission at 20 cm.
To look for O-type stars (likely responsibles of ionizing
the surrounding gas), we use the $J$, $H$, and $K$ apparent
magnitudes obtained from the 2MASS Point Source Catalog to derive 
the absolute {\it JHK} magnitudes. To perform that, we assume a distance of about 3.7 kpc and
we obtain the extinction for each source from the {\it (J-H)} and {\it (H-K)} colors.  We assume 
the interstellar reddening law of \citet{rieke85}~($A_J/A_V$=0.282; $A_H/A_V$=0.175 and
$A_K/A_V$=0.112) and the intrinsic colors $(J-H)_0$ and $(H-K)_0$
obtained  from \citet{mar06}.
By comparing the derived absolute magnitudes with those tabulated by
\citet{mar06}, we find that seven sources in Region 1, \#2, \#4, \#5,
\#6, \#11, \#12, \#14, and five sources in Region 2, \#16, \#17,
\#18, \#19 and \#20, have absolute {\it JHK} magnitudes that are in
agreement with an O-type star (see Table \ref{MSSC}).

Finally, taking into account that the exciting star
candidates are expected to be in a PAHs hole, we 
discard sources \#2, \#4 and \#11 as the responsible of generate G35.6. 
The rest of the exciting star candidates, sources
\#5, \#6, \#12 and \#14 are located in projection inside the radio
continuum and 24~$\mu$m emissions; among them, sources
\#6 and \#12 are located close to the maximum of the 24~$\mu$m emission
as could be expectable for an exciting star. On the other hand, as can be seen in Figure \ref{Ostars},
source \#6, is located into a hole of the 5.8~$\mu$m emission (see
zoom of the region in the figure). It is well known 
that the exciting star(s) of an HII region generate a cavity of dust
and gas through the action of the radiation pressure on the dust
grains \citep{gail79}, which suggests that source \#6 is the more likely candidate to
be the exciting star of G35.6. 
On the other hand, using the same assumptions that in R1, we found
that based on the radio continuum flux at 1.4 GHz in R2, the exciting
stars of NVSS 185938+020012 would be later than an O9.5V star. The later
spectral type stars found in R2 could be sources \#16 and \#18.

It would be very useful to have UBV fluxes to perform a better photometry in order to identify
without doubt the exciting stars, however it is very difficult to obtain these fluxes because of the 
interstellar absorption towards this region in the Galaxy.

\section{Star Formation}

In Sec. \ref{secmolec} we show that the HII region G35.6 is evolving and affecting a molecular cloud, presenting 
an excellent scenario to probe triggered star formation. In this section we look for young stellar objects (YSOs) around
G35.6. YSOs are generally classified according to their evolutionary stage: class I are the youngest sources, embedded in
dense envelopes of gas and dust, and class II are sources whose emission is originated mainly in the accretion disk
around the central protostar. In both cases, a YSO will show an infrared excess that cannot be attributed to 
the scattering and absorption of the ISM along the line of sight. On the contrary, this infrared excess is mainly due to 
the presence of the envelope 
and/or the disk of dust around the central protostar. In other words, YSOs are intrinsically red sources.

\begin{figure}[h]
\centering
\includegraphics[width=9cm]{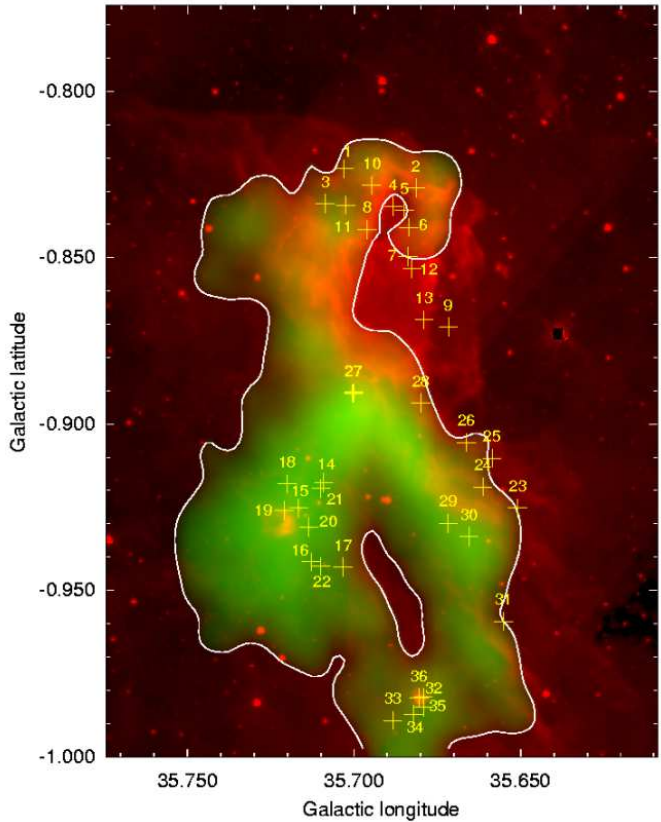}
\caption{Two-color image: the \3 emission integrated between 53 and 61 \ks is presented in green,
and the 8 $\mu$m emission, in red. For a better contrast, the \3 emission
scale is displayed in square root and bordered by a white contour. The yellow crosses indicate the position 
of the intrinsically red sources, i.e. sources satisfying the condition $m_{4.5}-m_{8.0}+\varepsilon\geq1$. We 
labeled the sources that appear to be related with the molecular gas around G35.6.}
\label{molec+YSO}
\end{figure}

\begin{table}
\caption{Near- and mid-IR fluxes of the sources satisfying the condition $m_{4.5}-m_{8.0}+\varepsilon\geq1$ around G35.6. 
The numbers of the YSO candidates correspond to the numeration in Fig. \ref{molec+YSO}}.
\label{yso_data}
\tiny
\begin{tabular}{ccccccccccc}
\hline\hline
Source & GLIMPSE Desig. & 2MASS Qual. & $J$ & $H$ & $K_{S}$ &  3.6 $\mu$m & 4.5 $\mu$m  &  5.8 $\mu$m &  8.0 $\mu$m & 24 $\mu$m \\
   &                &             & (mag) & (mag) & (mag) & (mag) & (mag) & (mag) & (mag) & (Jy) \\
\hline
YSO 1     & G035.7031-00.8232 & BAA  & 16.513 & 14.154 & 12.967 & 12.086 & 11.603 & 11.258 & 10.285 & \\
YSO 2     & G035.6814-00.8290 & AAA  & 15.830 & 13.625 & 12.495 & 11.284 & 10.692 & 10.049 & 9.182 & \\
YSO 3     & G035.7086-00.8337 & BAA  & 16.641 & 13.712 & 12.405 & 11.423 & 11.343 & 10.609 & 10.009 & \\
YSO 4     & G035.6882-00.8346 & UUA  & 18.264 & 15.119 & 14.434 & 12.445 & 11.992 & 11.545 & 10.877 & \\
YSO 5     & G035.6850-00.8358 & UAA  & 17.304 & 15.046 & 13.740 & 12.101 & 11.557 & 11.057 & 10.513 & \\
YSO 6     & G035.6836-00.8409 & UAE  & 18.038 & 15.025 & 13.485 & 11.546 & 11.054 & 10.487 & 9.958 & \\
YSO 7     & G035.6838-00.8496 & UAA  & 17.579 & 15.185 & 14.077 & 12.942 & 12.344 & 12.307 & 11.169 & \\
YSO 8     & G035.6961-00.8415 & N/A  &        &        &        & 13.176 & 11.297 &  9.674 &  8.558 & 0.14  \\
YSO 9   & G035.6716-00.8708 & AAA  & 15.493 & 13.459 & 12.624 & 11.955 & 11.845 & 11.431 & 10.973 &  \\
YSO 10   & G035.6947-00.8282  & N/A  &        &        &        & 13.616 & 13.762 &        &  9.994 &  \\
YSO 11   & G035.7025-00.8342 & N/A  &        &        &        & 14.065 & 13.908 &        &  9.828 & \\
YSO 12   & G035.6826-00.8533 & N/A  &        &        &        & 14.114 & 13.431 &        &  10.776 & \\
YSO 13   & G035.6790-00.8687 & N/A  &        &        &        & 13.970 & 13.690 &        &  10.811 & \\
YSO 14     & G035.7094-00.9176 & N/A  &      &       &          & 13.583 & 13.145 & 11.983 & 11.299 & \\
YSO 15     & G035.7166-00.9252 & AAA  & 15.776 & 13.607 & 12.112 & 10.110 & 9.568  & 8.992  & 8.213 & 0.05  \\
YSO 16    & G035.7127-00.9414 & AAA  & 15.713 & 14.113 & 13.148 & 11.797 & 11.196 & 10.718 & 10.142 & \\
YSO 17    & G035.7032-00.9430 & N/A  &      &       &          & 13.242 & 12.489 & 11.977 & 11.313  & \\
YSO 18    & G035.7201-00.9180 & N/A  &      &       &          & 14.267 & 15.532 & 12.876 & 11.596 &  \\
YSO 19   & G035.7209-00.9260 & UAA  & 18.218 & 14.992 & 13.379 & 10.983 &  9.856 &  9.109 &  8.363 &  0.16 \\
YSO 20   & G035.7135-00.9309 & UUA  & 18.257 & 16.988 & 13.519 & 11.184 & 10.301 &  9.740 &  9.235 &  0.05  \\
YSO 21   & G035.7099-00.9193 & N/A  &        &        &        &        & 13.787 & 12.354 &  11.781 & 0.03 \\
YSO 22   & G035.7103-00.9428 & UAA  & 16.441 & 14.637 & 13.419 & 12.754 & 12.716 & 11.904 &  11.944 & \\
YSO 23    & G035.6509-00.9252 & AAA  & 15.825 & 13.779 & 12.688 & 12.018 & 11.926 & 11.673 & 10.935 & \\
YSO 24   & G035.6613-00.9189 & UAA  & 17.288 & 14.442 & 13.061 & 12.238 & 11.995 & 11.687 & 11.049 & \\
YSO 25    & G035.6585-00.9103 & UAA  & 18.328 & 15.205 & 13.736 & 12.791 & 12.505 & 12.328 & 11.523 & \\
YSO 26    & G035.6663-00.9055 & UAA  & 17.509 & 13.978 & 12.464 & 11.240 & 11.208 & 10.885 & 10.344 & \\
YSO 27    & G035.7002-00.8907 & UUA  & 14.254 & 12.706 & 13.451 & 10.951 &  9.755 &  8.856 &  8.321 & 0.14 \\
YSO 28   & G035.6799-00.8936 & UAA  & 17.204 & 15.356 & 14.118 & 13.152 & 12.933 &        &  10.466 &  \\
YSO 29   & G035.6717-00.9298 & N/A  &        &        &        & 13.379 & 12.823 &        &  11.280 & \\
YSO 30   & G035.6655-00.9338 & N/A  &        &        &        & 14.952 & 13.166 & 11.493 &  11.337 & 0.04 \\
YSO 31    & G035.6551-00.9593 & AAA  & 16.315 & 14.103 & 13.002 & 12.365 & 12.317 & 12.050 & 11.254 & \\
YSO 32    & G035.6791-00.9821 & UAA  & 15.364 & 14.161 & 12.039 & 10.210 &  9.916 &  9.250 &  8.524 & \\
YSO 33   & G035.6883-00.9891 & BAA  & 16.607 & 14.877 & 14.225 & 13.284 & 13.150  &        &  11.873 & \\
YSO 34   & G035.6820-00.9873 & UAA  & 17.289 & 14.683 & 13.584 & 11.714 & 11.025 & 10.649 &  10.107 & \\
YSO 35   & G035.6790-00.9851 & UUB  & 18.284 & 15.686 & 14.581 & 11.914 & 10.942 & 10.260 &  9.520 & \\
YSO 36   & G035.6804-00.9823 & UAA  & 14.186 & 11.459 & 9.535 & 7.425 & 6.711 & 5.985 &  4.827 & \\
\hline
\end{tabular}
\tablefoot{2MASS Qual.: A and B are the best photometric qualities, with a SNR $\leq10$ and $\leq7$,
respectively. E means that the source magnitude is questionable. And U means that the magnitude value is an upper limit.}

\end{table}

\citet{robi08} defined a color criterion to identify intrinsically red sources using data from
the {\it Spitzer}-IRAC bands.
Intrinsically red sources satisfy the condition  $m_{4.5}-m_{8.0}\geq1$, where $m_{4.5}$ and $m_{8.0}$ are
the magnitudes in the 4.5 and 8.0 $\mu$m bands, respectively.
On the other hand, an externally reddened source is a source
not intrinsically red (such as main sequence stars) which appears red because of the interstellar effects. They
satisfy the condition $m_{4.5}-m_{8.0}<1$ and their spectral energy distributions (SEDs) are well fitted by stellar
photosphere models with interstellar extinction.
In order to consider the errors in the magnitudes, we use the following color criterion to select
intrinsically red sources: $m_{4.5}-m_{8.0}+\varepsilon\geq1$, where
$\varepsilon=\sqrt{(\Delta_{4.5})^2+(\Delta_{8.0})^2}$ and
$\Delta_{4.5}$ and $\Delta_{8.0}$ are the errors of the 4.5 and 8.0 $\mu$m bands, respectively.
On Fig. \ref{molec+YSO} we show the distribution of the sources extracted from the GLIMPSE catalog
around G35.6 that satisfy the previous criterion (we only considered sources with detections in both
4.5 and 8.0 $\mu$m bands).
The intrinsically red sources are distributed into four groups.
The first group (Group 1) is found towards the north and includes sources 1 to 13.
A second group (Group 2) of sources is located over the
southeastern molecular structure. This portion of the molecular cloud
is far from G35.6 and probably is not being perturbed by the HII region.
Sources 23, 24, 25, 26, 29 and 30 form Group 3 which appears in the molecular
gas likely associated with the border of the radio continuum source NVSS 185938+020012.
Then, we identify a fourth group (Group 4) towards the southern portion of the molecular cloud,
far from the HII regions. And finally, sources 27, 28 and 31 that are not part of any group.
Sources 27 and 28, taking into account their position, could be related to G35.6 southern
border.
On Table \ref{yso_data} we report the fluxes of the intrinsically red sources in the 2MASS
and {\it Spitzer}-IRAC bands, specifying the GLIMPSE designation (Col. 2) and the 2MASS photometric quality (Col. 3).
In the case of sources 8, 15, 19, 20, 21, 27, and 30, we obtained their fluxes at 24 $\mu$m from the MIPS image and
are presented in Col. 11 of the table.

\begin{figure}[h]
\centering
\includegraphics[width=9cm,angle=-90]{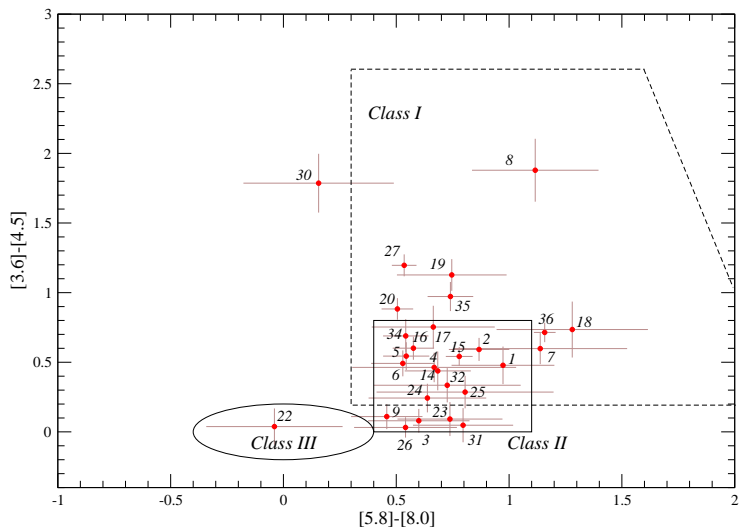}
\caption{Color-color diagram [5.8]-[8.0] versus [3.6]-[4.5] for sources of Table \ref{yso_data}
with detections in the four {\it Spitzer}-IRAC bands.
The regions indicate the stellar evolutionary stage based on the photometric criteria of \citet{all04}.}
\label{ccdiagr}
\end{figure}

So far we have identified the intrinsically red sources located right upon the molecular gas around
the HII region G35.6. However, according to \citet{robi08} intrinsically red sources may include YSOs, planetary 
nebulae (PNe), galaxies, AGNs, and AGB stars. We must apply an additional constraint to these sources 
in order to discern their real nature. 
Regarding extragalactic sources, \citet{robi08} pointed out that at most 0.4\% of the intrinsically red 
sources selected by the color criterion $m_{4.5}-m_{8.0}>1$ are galaxies and AGNs. 
So, there is a little probability to find an extragalactic source in our short sample of red sources.
Concerning to AGB stars, we look in catalogues for these stars. In the region analyzed in this work
there are not any AGB star catalogued.
To look for YSOs and PNe candidates we constructed a color-color (CC) diagram [5.8]-[8.0] versus [3.6]-[4.5] 
with the sources from Table \ref{yso_data} that have flux detections in the four {\it Spitzer}-IRAC bands. 
We used the photometric criteria of \citet{all04} to identify class I and II YSOs (Fig. \ref{ccdiagr}).    
From its positions in this color-color diagram we find that only source 22 cannot be classified
as a class I or II YSO. Source 30 falls outside the class I region but if we take into account the errors in 
the fluxes we should consider it as a YSO candidate. On the other hand, sources 7 and 18 are found in the 
CC diagram close to the location of PNe, which typically have [5.8]-[8.0]$>1.4$ \citep{cohen07}.
However, we will consider them as YSO candidates too. 

Finally, we fitted the spectral energy distribution (SED) of the sources from which we obtained fluxes at 
24 $\mu$m from the MIPS image using the tool developed by \citet{robi07} and available
online\footnote{http://caravan.astro.wisc.edu/protostars/}.
We assume an interstellar absorption between 12 and 22 mag.
These values where obtained from the 2MASS {\it J-H} versus {\it H-Ks} color-color diagram (not presented here) 
constructed with
the sources with the best photometric quality (AAA) within a circle of~8\m~in radius centered at G35.6.
The lower value is compatible with the expected extinction towards star forming regions which, according to
\citet{neckel80}, is generally greater than 10 magnitudes. The upper value is in agreement with the 
visual absorption of $A_{v}\sim20$~mag obtained from $A_{v}=5\times{10^{-22}}N(H)$ \citep{bohlin78}, 
where N(H) $=$ N(HI)$+$2N(H$_{2}$) is the line-of-sight hydrogen column density towards this region, which
is about 4 $\times 10^{22}$ cm$^{-2}$. This value was obtained from the HI column density derived from the VGPS HI 
data and from the H$_{2}$ column density derived from the \3 J=1--0 data.

The SED good fitting models are selected according to the following condition:
$\chi^{2} - \chi^{2}_{YSO} < 2N$,
where $\chi^{2}_{YSO}$~is the $\chi^{2}$~of the YSO best-fit model,
and {\it N} is the number of input data fluxes (fluxes specified as upper limit do not contribute
to {\it N}). Hereafter, we refer to models satisfying the above equation as ``selected models''.
The fitting tool also fits the data to a stellar photosphere and defines the parameter $\chi^{2}_{\star}$
to evaluate the goodness of the fitting. By making a comparison between $\chi^{2}_{YSO}$ and $\chi^{2}_{\star}$  
we can confirm which sources are in fact YSOs and which sources may be stars externally reddened by the ISM. 
The SED fitting allows us to establish the evolutionary stage of the YSO candidates 
by considering the physical parameters of the sources:
the central source mass $M_\star$, the disk mass $M_{disk}$,
the envelope mass $M_{env}$, and the envelope accretion rate $\dot{M}_{env}$. According to \citet{robi06},
stage I YSOs are those that have $\dot{M}_{env}/M_\star>10^{-6}~yr^{-1}$, i.e., protostars with
large accretion envelopes; stage II are those with $M_{disk}/M_\star>10^{-6}$ and
$\dot{M}_{env}/M_\star<10^{-6}~yr^{-1}$, i.e., young objects with prominent disks;
and stage III are those with $M_{disk}/M_\star<10^{-6}$ and
$\dot{M}_{env}/M_\star<10^{-6}~yr^{-1}$, i.e., evolved sources where the flux is dominated by the central source.

\begin{table}[h]
\caption{Parameters derived from the SED fitting of sources from which we obtained fluxes at 24 $\mu$m from the MIPS image.}
\label{tablesed}
\centering
\tiny
\begin{tabular}{ccccccccc}
\hline\hline
Source& $\chi^{2}_{YSO}/{\it N}$ & $\chi^{2}_{\star}/{\it N}$ &{\it n} &M$_{\star}$ &M$_{disk}$ &M$_{env}$ & 
$\dot{M}_{env}$&Stage  \\
           &       &  &           & (M$_{\odot}$) & (M$_{\odot}$) & (M$_{\odot}$) & (M$_{\odot}/yr$)&  \\
\hline
YSO 8&1&520&12&$1 - 4$&$1\times{10^{-3}} - 2\times{10^{-1}}$&$5\times{10^{-8}} - 3\times{10^{1}}$&$0 - 3\times{10^{-4}}$&I \\
YSO 15&0.4&50&109&$3 - 5$&$1\times{10^{-5}} - 1\times{10^{-1}}$ & $6\times{10^{-9}} - 2$&$0 - 2\times{10^{-7}}$& II\\
YSO 19&1&$>9999$&7&$4$&$3\times{10^{-3}} - 5\times{10^{-2}}$&$2\times{10^{-4}}-8\times{10^{-2}}$ & $0 - 9\times{10^{-6}}$ & I, II \\
YSO 20&28&$>9999$&1&5&$3\times{10^{-4}}$ & $4\times{10^{-6}}$ & 0 & II \\
YSO 21&0.1&505&72&$0.2 - 5$&$6\times{10^{-4}} - 3\times{10^{-1}}$&$9\times{10^{-3}} - 2\times{10^{2}}$&$2\times{10^{-6}} - 9\times{10^{-4}} $ & I \\
YSO 30&1.5&580&12&$2 - 13$&$2\times{10^{-3}} - 5\times{10^{-1}}$&$2 - 5\times{10^{2}}$&$5\times{10^{-5}} - 6\times{10^{-3}}$ & I \\
YSO 27&1.3&$>$9999&2&$8 - 18$&$0 - 4\times{10^{-2}}$&$7\times{10^{1}} - 2\times{10^{2}}$&$8\times{10^{-5}} - 2\times{10^{-3}}$& I \\
\hline
\end{tabular}
\end{table}

\begin{figure}[h]
\centering
\includegraphics[width=12cm]{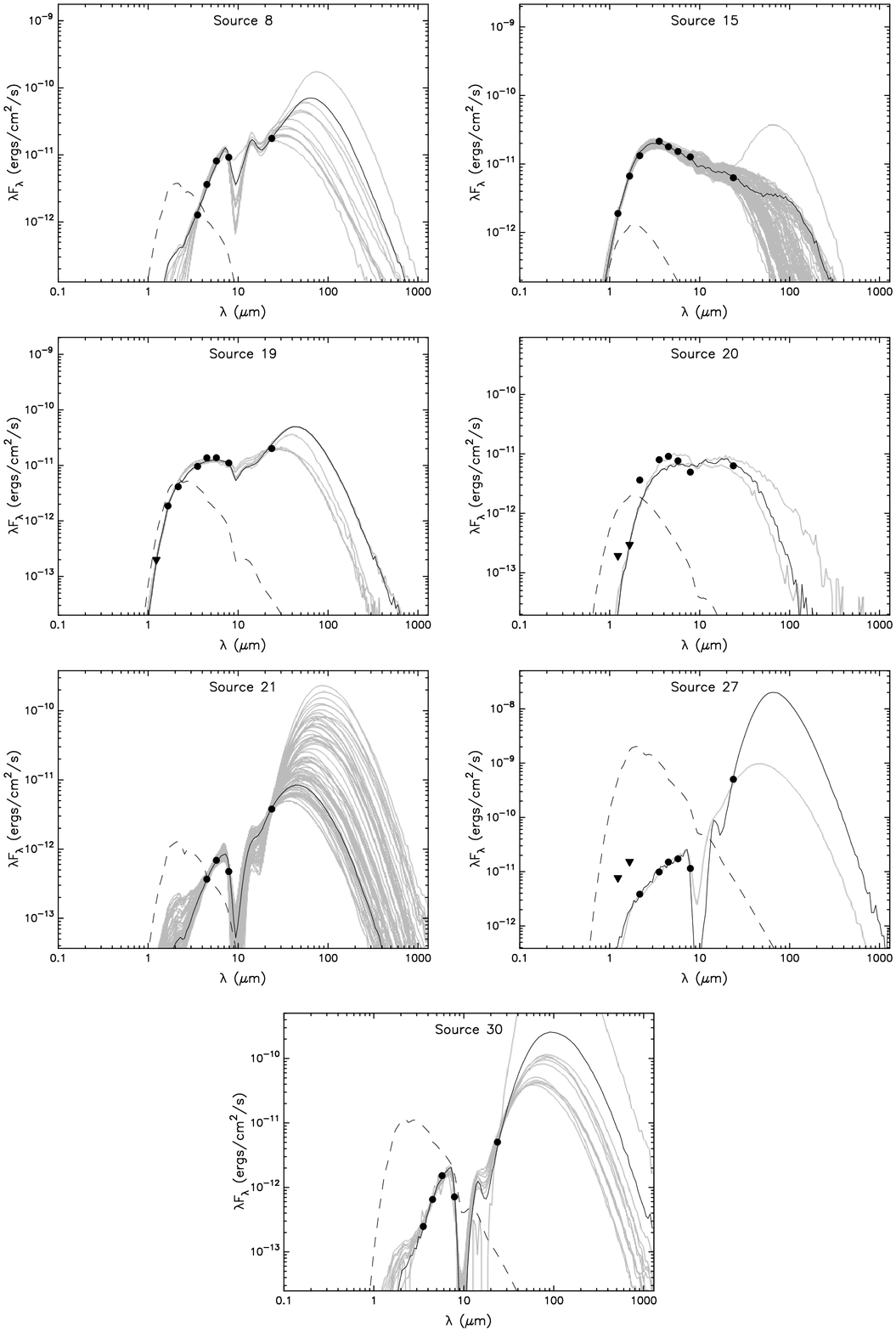}
\caption{SED of sources from which we obtained fluxes at 24 $\mu$m from the MIPS image. 
The sources are numbered according to Table \ref{tablesed} and Figs. \ref{molec+YSO} and \ref{ccdiagr}. 
In each panel, black line shows the best fit, and the 
gray lines show subsequent good fits. The dashed line shows the stellar photosphere corresponding to the central 
source of the best fitting model, as it would look in the absence of circumstellar dust. The points are the 
input fluxes.}
\label{sedFit}
\end{figure}

In Table \ref{tablesed}, we report the main results of the fitting output for the YSO candidates 
from which we obtained fluxes at 24 $\mu$m from the MIPS image. In Col. 2 and 3 we report 
the $\chi^{2}$ per data point of the YSO and stellar photosphere best-fit model, respectively, 
and in Col. 4 the number of models satisfying the $\chi^{2}$ equation.
The remaining columns
report the physical parameters of the source, specifying the range of values of the selected models:
central source mass, disk mass, envelope mass, and envelope accretion rate, respectively. Following the criteria of
\citet{robi06}, the last column indicates the evolutionary stage inferred from the inspection of the selected models.
Figure \ref{sedFit} shows the SED of these sources.
From this analysis can be appreciated that for source 15 the flux at the longer wavelengths comes mainly from
the disk, indicating that it is a class II YSO, in coincidence with its
position in the Spitzer-IRAC CC diagram. The SED for source 20 also shows
the characteristics of a class II YSO.
In the case of source 19 the selected models indicate that this source could be 
stage I and II.
For sources 8, 21, 27, and 30 the selected models are stage I and the
SED shows that the flux at the longer wavelengths is dominated by the
envelope flux. These sources, together with source 19, are located in the region of class I YSO
in the CC diagram (except for source 21 that lacks flux at 3.6 $\mu$m),
confirming their youth.

From Fig. \ref{ccdiagr} and the SED analysis we can confirm the presence of 
YSOs around G35.6. 
Thus, we conclude that the region is indeed active in star 
formation and we suggest that the birth of some of the YSOs, mainly those belonging to Group 1 and sources 27 and 28
could have been triggered by the expansion of the HII region G35.6. Most of the remaining intrinsically red sources
belonging to Group 2, 3, and 4 may also be YSOs but their position far from the HII region does not allow us to
confirm that their formation was triggered by G35.6.

\section{Collect and collapse scenario}

In order to determine if the collect and collapse mechanism is the
responsible for the star formation that is taking place in the
periphery of the HII region G35.6, we estimate and compare
the age of the HII region and the fragmentation time predicted by the
theoretical models of \citet{whit94a,whit94b}.

Using a simple model described by \citet{dys80} we
calculate the age of the HII region at a given radius $R$ as:
$$t(R)=\frac{4~R_s}{7~c_s}\left[\left(\frac{R}{R_s}\right)^{7/4}-1\right],$$
where $c_s$ is the sound velocity in the ionized gas ($c_s$=10 km
s$^{-1}$) and $R_s$ is the radius of the Str\"omgren sphere, given
by $R_s=(3N_{\rm uv}/4\pi n_0^2\alpha_B)^{1/3}$, where $\alpha_B$=2.6
$\times 10^{-13}$ cm$^3$ s$^{-1}$ is the hydrogen recombination
coefficient to all levels above the ground level. $N_{\rm uv}$ is the
total number of ionizing photons per unit of time emitted by the
star(s), and $n_0$ is the original ambient density.

Taking into account the results of Sec. 6, we consider a Lyman
continuum photon flux of 1.0 $\times 10^{48}$ ph s$^{-1}$.  Adopting
a radius of $\sim$ 1.5$\arcmin$ for the HII region, a distance of 3.7
kpc, and an original ambient density of $\sim (1\pm0.5)\times 10^{3}$~cm$^{-3}$, we
derive a dynamical age between 0.18 and 0.35 Myr for G35.6. To coarsely estimate the
original ambient density (assuming an error of 50\%) we distributed the above calculated mass of
the molecular shell, $\sim$ 10$^4$ M$_{\odot}$, over an ellipsoid of
revolution with semiaxes of 3 and 7~pc that encloses the molecular and
ionized gas. 

As analyzed in Sec. \ref{secmolec}, the morphology of the molecular
gas that encircles the HII region suggests that the expansion of G35.6
is collecting the gas at its periphery. Finally, we ask if the
fragmentation of the collected layer of material can be taking place
in the region. To answer it, we estimate when the fragmentation of the
collected layer should occur according to the Whitworth's models
Assuming a turbulent velocity in the
collected layer a$_s$ ranging between 0.2 and 0.6 km s$^{-1}$ \citep{whit94b}, a Lyman
continuum photon flux of 1.0 $\times 10^{48}$ ph s$^{-1}$ , and the
previously estimated original ambient density of $\sim (1\pm0.5)\times 10^{3}$~cm$^{-3}$, 
we find that the fragmentation process in the
periphery of G35.6 should occur between 1.6 and 5.3 Myr after its
formation, a time larger than the G35.6 dynamical age derived above.
The range for the fragmentation time arises from considering the error in 
the original ambient density and the range in the turbulent velocity.
Thus, we conclude that the formation of the YSOs lying at the border
of the HII region most probably results from other processes, such as
the radiative driven implosion (RDI) mechanism, which consists in
interactions of the ionization front with pre-existing condensations
\citep{lefloch94}, or small-scale Jeans gravitational instabilities
in the collected layer.

\section{Summary}

Using multiwavelength surveys and archival data, we studied the ISM towards the HII region
G35.673-00.847 (G35.6). This work is part of a systematic study that we are performing with the aim to increase
the observational evidence of triggered star formation in the surroundings of HII regions.
The main results can be summarized as follows:

(a) The PAH emission around G35.6 seen at 8 $\mu$m~shows that the HII region has an almost semi-ring like shape 
with a cut towards the galactic west. The radius of this semi-ring is about 1\farcm5.
The 24 $\mu$m emission reveals the presence of hot dust in the interior of 
the HII region. 

(b) The radio continuum emission shows that towards the south of G35.6, also identified as NVSS 185929$+$020334,
lies the radio source NVSS 185938$+$020012, probably 
another HII region. From the HI absorption analysis we conclude that both sources are located at the same
distance, and from the central velocity of the related molecular gas, we estimate that the whole complex 
is at the kinematic distance of  $\sim 3.7$ kpc.

(c) Using the \3 J $=$ 1--0 transition we analyze the molecular gas around G35.6. We find a molecular shell 
composed by clumps distributed around the HII region, suggesting that its expansion is collecting the material. 
The molecular shell has a density of about 10$^{4}$ cm$^{-3}$.

(d) From a photometric study and a SED analysis we find several sources (YSO candidates) very likely embedded in 
the molecular shell.

(e) We give some indirect evidence suggesting the possible location and properties of the exciting star(s) of
G35.6 and NVSS 185938$+$020012. In the case of G35.6, from the radio continuum flux, the near-IR photometry and 
the physical location of the analyzed sources, we find four candidates, likely O-type stars, to be the ionizing 
agent of the HII region. 
Among them, two are located close to the maximum of the 24 $\mu$m emission, and one of them (our source \#6) appears
into a hole of 5.8 $\mu$m emission, suggesting to be the most likely candidate. In the case of 
NVSS 185938$+$020012, we suggest that the exciting star(s) would be later than an O9.5V star.

(f) Analyzing the HII region G35.6 dynamical age and the fragmentation time of the molecular shell surrounding
the HII region, we discard the collect and collapse as the mechanism responsible for the YSOs formation. 
We propose other possible processes of formation, such as radiative driven implosion and/or small-scale Jeans 
gravitational instabilities in the collected layer.

\section*{Acknowledgments}

We wish to thank the anonymous referee whose comments and suggestions have helped to considerably improve the paper.
S.P. is member of the {\sl Carrera del 
investigador cient\'\i fico} of CONICET, Argentina. A.P. and M.O. are doctoral and postdoctoral fellows of CONICET, 
Argentina, respectively. 
This work was partially supported by Argentina grants awarded by UBA, CONICET and ANPCYT.

\bibliographystyle{aa}  
\bibliography{biblio}
\IfFileExists{\jobname.bbl}{}
{\typeout{}
\typeout{****************************************************}
\typeout{****************************************************}
\typeout{** Please run "bibtex \jobname" to optain}
\typeout{** the bibliography and then re-run LaTeX}
\typeout{** twice to fix the references!}
\typeout{****************************************************}
\typeout{****************************************************}
\typeout{}
}

\label{lastpage}
\end{document}